\documentclass{aastex63}

\graphicspath{{./}}

\usepackage{amsmath}
\usepackage{float}

\begin{document}

\title{Abundance of LIGO/Virgo Black Holes from Microlensing Observations of Quasars with Reverberation Mapping Size Estimates}

\author{A. ESTEBAN-GUTI\'ERREZ} 
\affiliation{Instituto de Astrof\'{\i}sica de Canarias, V\'{\i}a L\'actea S/N, La Laguna 38205, Tenerife, Spain}
\affiliation{Departamento de Astrof\'{\i}sica, Universidad de la Laguna, La Laguna 38206, Tenerife, Spain}
\author{N. AG\"UES-PASZKOWSKY}
\affiliation{Instituto de Astrof\'{\i}sica de Canarias, V\'{\i}a L\'actea S/N, La Laguna 38205, Tenerife, Spain}
\affiliation{Departamento de Astrof\'{\i}sica, Universidad de la Laguna, La Laguna 38206, Tenerife, Spain}
\author{E. MEDIAVILLA}
\affiliation{Instituto de Astrof\'{\i}sica de Canarias, V\'{\i}a L\'actea S/N, La Laguna 38205, Tenerife, Spain}
\affiliation{Departamento de Astrof\'{\i}sica, Universidad de la Laguna, La Laguna 38206, Tenerife, Spain}
\author{J. JIM\'ENEZ-VICENTE}
\affiliation{Departamento de F\'{\i}sica Te\'orica y del Cosmos, Universidad de Granada, Campus de Fuentenueva, 18071 Granada, Spain}
\affiliation{Instituto Carlos I de F\'{\i}sica Te\'orica y Computacional, Universidad de Granada, 18071 Granada, Spain}
\author{J. A. MU\~NOZ}
\affiliation{Departamento de Astronom\'{\i}a y Astrof\'{\i}sica, Universidad de Valencia, 46100 Burjassot, Valencia, Spain.}
\affiliation{Observatorio Astron\'omico, Universidad de Valencia, E-46980 Paterna, Valencia, Spain}
\author{S. HEYDENREICH}
\affiliation{Argelander-Institut f\"ur Astronomie, Auf dem H\"ugel 71, 53121, Bonn, Germany}

\begin{abstract}

Assuming a population of Black Holes (BHs) with masses in the range inferred by LIGO/Virgo from BH mergers, we use quasar microlensing observations to estimate their abundances.  We consider a mixed population of stars and BHs and the presence of a smooth dark matter component. We adopt reverberation mapping estimates of the quasar size. According to a Bayesian analysis of the measured microlensing magnifications, a population of BHs with masses $\sim$ 30$M_{\odot}$ constitutes less than 0.4\% of the total matter at 68\% confidence level (less than 0.9\% at 90\% confidence). We have explored the whole mass range of LIGO/Virgo BHs finding that this upper limit ranges from 0.5\% to 0.4\% at 68\% C.L. (from 1.1\% to 0.9\% at 90\% C.L.) when the BHs mass change from 10 to 60$M_{\odot}$. We estimate a 16\% contribution from the stars, in agreement with previous studies based on a single mass population that do not consider explicitly the presence of BHs. These results are consistent with the estimates of BH abundances from the statistics of LIGO/Virgo mergers and rule out that PBHs (or any type of compact objects), in this mass range constitute a significant fraction of the dark matter.

\end{abstract}


\section{Introduction \label{intro}}

According to the results from LIGO/Virgo collaboration (cf. GWTC-1 and GWTC-2 by Abbott et al. 2019a and Abbott et al. 2020a) the masses involved in the binary black hole (BBH) mergers detected from gravitational wave observations, can be significantly larger than originally expected for BHs of stellar origin (a typical value of 30$\, M_\odot$, but with estimates as large as 60$\, M_\odot$). This fact, and the low effective spins of the components have led to speculation that some of these BHs could be of primordial origin, and even that these Primordial Black Holes (PBHs) of intermediate mass (20$-$200 $M_{\odot}$) could constitute a substantial part of the dark matter in the Universe (see, e.g.,  Carr \& K{\"u}hnel, 2020). \\

Several paths for the stellar formation of these BBHs have now been proposed with specific conditions and/or physical processes involved, including common envelope, chemical homogeneous scenario and dynamical evolution (see extensive list in Abbott et al. 2020b) and new limits have been set on the abundance of PBHs on different grounds, cooling down the initial excitement. However, the recent discovery of some events like GW180914 (Abott et al. 2020a) with a BH in the mass gap between neutron stars and BHs, and GW190521 (Abbott et al. 2020b) in the mass gap predicted by the pair-instability supernovae theory has again reopened the possibility that some of the BBHs have a non-stellar origin and, if so, they could be more abundant than previously thought. \\

Complementary to the detection rate of binary mergers via gravitational waves, an alternative method to estimate the abundance of BHs but also of any type of compact objects (including stars) is quasar microlensing (Chang \& Refsdal 1979; Wambsganss 2006). When a quasar is multiply imaged by an intervening galaxy (the lens), the granulation of the matter of the lens galaxy in compact objects (microlenses) can affect the gravitational potential inducing changes in the flux of the lensed quasar images with respect to the ones expected if the matter in the galaxy were smoothly distributed. This effect is sensitive to both, the mass and the abundance of any population of compact objects of the lens galaxy (see, e.g., Schechter \& Wambsganss 2004, Mediavilla et al. 2009; Pooley et al. 2012; Schechter et al. 2014; Jim{\'e}nez-Vicente et al. 2015a, 2015b, Mediavilla et al. 2017, Schechter 2018, Jim{\'e}nez-Vicente \& Mediavilla, 2019, Esteban-Guti{\'e}rrez et al. 2020). \\

To limit the abundance of BHs, a mixed population of microlenses including both stars and BHs, needs to be considered\footnote{A more generic, qualitative approach, based on the different impact of the finite size of quasars on the contribution of stars or BHs to microlensing flux magnification is presented in an accompanying letter.}. This involves many unknowns (at least four: the masses and abundances of both components), which makes this study difficult. Previous works circumvented this problem with indirect approaches related to the reinterpretation of results based on a single-mass microlens population (Mediavilla et al. 2017) or to generic studies of the impact of a bimodal distribution in the statistics of microlensing magnifications (Esteban-Guti{\'e}rrez et al. 2020). None of these works support the existence of a significant population of intermediate mass BHs, but they are indirect, qualitative {and incomplete (Esteban-Guti{\'e}rrez et al. 2020 do not include the smooth matter component). On the other hand, the theoretical approach of Esteban-Guti{\'e}rrez et al. (2020) requires to work in the high spatial resolution limit, small as compared with quasar disk sizes inferred from reverberation mapping (RM) determinations (cf. Mediavilla et al. 2017 and references therein)}. \\

Now, from LIGO/Virgo results, we have estimates of the mass of the merging BHs which can be used to remove one of the unknowns of the problem (e.g. Abbott et al. 2020b). If we make an educated guess about the mass of the stars, then we are left with only two main parameters of interest, the abundances of stars and BHs, and the problem becomes tractable even with a direct approach. Thus, the main objective of this work is to estimate at once the most likely abundances of both stars and BHs from a Bayesian study of the observed microlensing magnifications of the images of lensed quasars. Apart from the mixed population of stars and BHs, we also consider a smooth dark matter component and a quasar source size in agreement with current RM estimates. In principle, each one of the components,  BHs and stars, could have its own mass distribution around a central value. As far as these distributions are smooth, the main factor affecting microlensing is the high ratio between the mean masses of stars and BHs, and, as we will address later, a bimodal mass spectrum with a unique type of stars and a unique type of BHs is a reasonable approximation to simulate the impact of a mixed population (cf. Schechter et al. 2004). \\

The article is organized as follows. In \S2 we describe the methodology and data used. In \S3 we present the results of the application of the bimodal mass-spectrum model to quasar microlensing data. In \S4 we discuss the robustness of the results and their implications in the context of LIGO/Virgo discoveries and the existence of a significant number of PBHs. Finally, in this same section (\S4) we summarize the conclusions. 
\\

\section{Methods and data \label{methods}}

To describe the population of microlenses, we consider a bimodal distribution of BHs and stars. For the mass of the stars we adopt $m_{stars}=0.2 \rm\, M_{\odot}$, as representative of the mean stellar mass. Although the mass function of the population of BHs in binaries detected by LIGO/Virgo is not {fully} determined, the detections are compatible with a smooth mass function between (roughly) 4 and 80 $M_{\odot}$ with an average mass of 25-30 $M_{\odot}$ (cf. Abbott et al. 2019b, Abbott et al. 2020b). Therefore, despite the specific structure of the BH mass function (cf. Abbott et al. 2020b), 30 $M_{\odot}$ represents a characteristic mass of the merging BHs. In these circumstances, we can assume that microlensing statistics are not very sensitive to the specific mass function (see discussion in \S \ref{discussion}) and that we can model the microlensing effects of the BHs by a single mass distribution with a mass of $\sim 30M_{\odot}$. \\

In addition to stars and BHs, we also consider a smooth matter distribution which contributes to the total (projected) mass with a fraction $\alpha_{smooth}$. The free parameters in our model are the fraction of mass in BHs, ${\alpha_{BH}}_k=\{0,0.00625,0.0125, 0.025, 0.05, 0.1,0.2,0.4,0.8\}$,  and the fraction of mass in stars, ${\alpha_{stars}}_l=\{0,0.05, 0.1,0.2,0.3,0.4,0.5\}$. By definition, $\alpha_{stars}+\alpha_{BH}+\alpha_{smooth}=1$. \\

To evaluate the likelihood of the different values of $\alpha_{BH}$ and $\alpha_{stars}$, we use the microlensing data presented in Jim\'enez-Vicente et al. (2015a), consisting of a sample of 27 quasar image pairs seen through 19 lens galaxies, increased with 7 new image pairs (with one or more epochs depending on the image pair) and 6 new measurements (see Table 1). As explained in Mediavilla et al. (2009) (see their Equations 3 and 4), the microlensing magnification data presented in this work are really differences of microlensing magnification between two images of a same lensed quasar, $\Delta m_{ij}=m_{i}-m_{j}$.  \\

The first step in the simulations is, then, to estimate the likelihood of measuring a microlensing magnification, $m_i$, in one image of a multiple imaged quasar characterized\footnote{Each multiple imaged lens system is modeled with a SIS+$\gamma_e$ and each image of the system characterized by the dimensionless projected mass density (i.e., the convergence), $\kappa$, and shear, $\gamma$, at its location (see Mediavilla et al. 2009).} by its convergence, $\kappa_i$, and shear, $\gamma_i$, given a pair of values of the fractions of mass in BHs and stars (${\alpha_{BH}}_k$,${\alpha_{stars}}_l$): $p_{ikl}( m_i|{\alpha_{BH}}_k,{\alpha_{stars}}_l)$. Taking into account the uncertainties in the macrolens models,  to reduce the computation time we have considered only 9 values  for the convergence and shear, $\kappa (=\gamma)=\{0.25,0.35,0.45,0.55,0.65,0.75,0.85,0.95,1.05\}$. For each image, we take the ($\kappa,\gamma$) pair closest to the SIS+$\gamma_e$ values determined in Mediavilla et al. (2009) (see Table 1 for the new image-pairs). Only for 7 of the images is the distance in the ($\kappa,\gamma$) plane greater than 0.1 to one of the calculated models. In any case, we have repeated the calculations excluding these 7 images, obtaining negligible differences. We compute microlensing magnification maps for each lensed image in the range of values $\{ {\alpha_{BH}}_k\}\times \{ {\alpha_{stars}}_l\}$ and for each pair of values ($\kappa_i,\gamma_i$), for a total of $9\times9\times 7 = 567$ different models. The magnification maps are calculated using the inverse polygon mapping algorithm (see Mediavilla et al. 2006, 2011). To reduce the noise induced by the sample variance caused by the random realizations of the microlenses positions, we generate and average the PDFs of the 100 microlensing magnification maps calculated for each model (for a total of 56700 maps). Maps are 424 $\times$ 424 pixels in size, with a pixel size of 2 light-days\footnote{We have tested that this pixel size is small enough, by checking that it produces identical magnification histograms than those obtained from maps with smaller pixel size (0.5 and 1 light-days), when convolved with a source of $5$ light-days of size.} (of which 50 pixels are removed on each side to avoid border effects). Each magnification map is convolved with a Gaussian of sigma $r_s=5$ light-days, a typical quasar size according to RM studies (see Edelson et al. 2015, Fausnaugh et al. 2016, Jiang et al. 2017 and the discussion in Mediavilla et al. 2017), and normalized to the fiducial mean value of the magnification map. The normalized histogram of the magnification map is the PDF of the microlensing magnification, $p_{ikl}(m_i|{\alpha_{BH}}_k,{\alpha_{stars}}_l)$. 

The probability of measuring a differential microlensing magnification between images i and j, $\Delta m_{ij}$, is given by the cross-correlation of the single image probabilities (see, e.g.,  Equation 5 of Mediavilla et al. 2009), $p_{ijkl}( \Delta m_{ij}|{\alpha_{BH}}_k,{\alpha_{stars}}_l)=p_{jkl}(m_i|{\alpha_{BH}}_k,{\alpha_{stars}}_l)\star p_{ikl}( m_j|{\alpha_{BH}}_k,{\alpha_{stars}}_l)$. Thus, to estimate quantitatively the joint probability of the abundance of BHs and stars, $p(\alpha_{BH},\alpha_{stars}|\{\Delta m_{ij}\})$, given the observed microlensing magnifications, $\{\Delta m_{ij}\}$, we apply Bayes Theorem to each image pair, 
 \begin{equation}
 p_{klij}({\alpha_{BH}}_k,{\alpha_{stars}}_l|\Delta m_{ij})\propto p_{ijkl}( \Delta m_{ij}|{\alpha_{BH}}_k,{\alpha_{stars}}_l),
 \end{equation} 
 and obtain the total probability as the product of all the image pair individual probabilities,  $p_{kl}({\alpha_{BH}}_k,{\alpha_{stars}}_l|\{\Delta m_{ij}\})\propto \prod_{ij}{p_{klij}({\alpha_{BH}}_k,{\alpha_{stars}}_l|\Delta m_{ij})}$. \\

 To prove the ability of our Bayesian method to reproduce a known result, we generate 100 random samples of 44 mock measurements (corresponding to each one of the considered image pair measurements) for the cases $\alpha_{BH}=0$ and $\alpha_{BH}=0.025$ (with $\alpha_{stars}=0.1$). Then, we apply the analysis described above to the mock data. Figure \ref{1d_mock} shows the marginalized PDFs which recover very well the fiducial values with reasonable low scatter: $\alpha^{Bayes}_{stars}=0.11_{-0.06}^{+0.05}$, $\alpha^{Bayes}_{BH}<0.006$ at $68\%$ confidence level and $\alpha^{Bayes}_{stars}=0.14_{-0.10}^{+0.10}$, $\alpha^{Bayes}_{BH}=0.035_{-0.025}^{+0.022}$ at $68\%$ confidence level, for $\alpha_{BH}=0$ and $\alpha_{BH}=0.025$, respectively.
\\

\section{Results \label{results}}
 
 The resulting 2D PDF and the 1D marginalized PDFs in both $\alpha_{BH}$ and $\alpha_{stars}$ when we apply the procedure described in \S2 to the real microlensing data, are shown in Figure \ref{corner_plot_med}. According to this Figure, the probability of significant abundances of BHs is negligible. The 2D joint PDF shows a maximum at $\sim$20\% of stars with zero contribution from the BHs. This is confirmed by the marginalized 1D PDF of BHs, which peaks at zero. We find an expected value for the abundance of stars, $\alpha_{stars}=0.16_{-0.07}^{+0.06}$ at 68\% confidence level ($\alpha_{stars}=0.16_{-0.09}^{+0.12}$ at 90\%) in agreement with previous estimates (cf. Jim{\'e}nez-Vicente et al. 2015a, 2015b). For the BHs, we find an upper limit $\alpha_{BH} < 0.004$ at 68\% confidence level ($\alpha_{BH} < 0.009$ at 90\%). The strong constraint imposed on the BHs abundance is explained by the large number (44) of measurements considered (see Appendix A). \\
 
 To explore all the mass range inferred from LIGO/Virgo observations, we have repeated the calculations for 10 and 60$M_{\odot}$ respectively. Our results show that the upper limits for the abundance of BHs have a slight mass dependence with the BH mass, with an increase by roughly 20\% for $10 M_{\odot}$ and no evidence of decrease for the highest mass of $60 M_{\odot}$. Upper limits at 68\% (90\%) confidence level for the abundance of BHs move in the range of 0.5\%-0.4\% (1.1\%-0.9\%) for this mass range. \\

 It may be argued that the sample from Mediavilla et al. (2009) lacks high magnification microlensing events. We have repeated the calculations for the lens sample of Pooley et al. (2007), which includes several objects of high microlensing magnification\footnote{We have eliminated Q2237+0305 from their sample in this comparison, as this system is produced by a nearby lens, and has nearly 100\% of the mass density in form of compact objects.}. The baseline used to derive the microlensing magnifications by Pooley et al. (2007) is defined from the macrolens model. As models are subject to uncertainties and extinction can also be playing a role, microlensing magnifications computed using this baseline could be biased, hence favoring the BHs hypothesis. However, an important advantage of the model based baseline is that it is not subject to microlensing induced by compact objects of very large masses. It is for this reason that we have preferred to keep both samples separated. In Figure \ref{corner_plot_pooley} we present the PDF 2D and the 1D marginalized PDFs corresponding to Pooley et al. (2007) data, which are in good agreement with the results based in emission line flux measurements (see Figure \ref{corner_plot_med}). For
 this sample we also obtain a very low upper limit $\alpha_{BH} < 0.02$ at 68\% confidence level ($\alpha_{BH} < 0.04$ at 90\%). Among the quasars in Pooley et al. (2007) sample, there is SDSS J0924+0219, a system with a very strong demagnification for its D image, which, in principle, seems difficult to explain under the hypothesis of a population of stellar mass microlenses. However, Schechter \& Wambsganss (2002) already showed that saddle point images of relatively high macro magnification with a fraction of $\sim$80\% of smooth matter can easily produce  demagnifications of 2.5 magnitudes, as in this case. We have therefore carefully modeled this system, including a smaller than average size, in order to better reproduce its extreme demagnification. \\
 
 Finally, it is interesting to mention that our Figures \ref{corner_plot_med} and \ref{corner_plot_pooley} can also be useful to study the relative abundance of stars and BHs when the smooth component of dark matter is the quantity fixed a priori. Regions of constant smooth dark matter will be represented as straight lines ($\alpha_{stars}+\alpha_{BH}=1-\alpha_{smooth}$) of slope -1 and intercept of $1-\alpha_{smooth}$. We show some of these lines of constant $\alpha_{smooth}$ in both Figures \ref{corner_plot_med} and \ref{corner_plot_pooley}. Notice that the probability of a significant abundance of PMBHs is very small along any of these lines, although the constrast is lower for small values of $\alpha_{smooth}$. 
\\

\section{Discussion and Conclusions\label{discussion}}

We analyze, in first place,  the impact in the results of the modeling assumptions made in the calculations. We have considered a bimodal mass-spectrum, combination of two "monochromatic" distributions (i.e. of a single mass) for stars and BHs. The use of this simplified mass function is based on previous results from microlensing simulations. Schechter et al. (2004) show that the shape of the mass function of the microlenses is only important for markedly bimodal distributions with a large and comparable contribution to the mass density from microlenses of very different masses. Otherwise, for more realistic smooth stellar mass functions, the relevant parameter is the mean mass (specifically, the Geometric Mean, GM, according to Jim\'enez-Vicente \& Mediavilla 2019 and Esteban-Guti{\'e}rrez et al. 2020). Thus, considering a smooth mass-function for the stars around their mean mass would likely have a very limited impact in the results. However, the BHs mass range exceeds very much the one covered by the stars and, in fact, our analysis shows that a slight anti-correlation of the BHs abundance with mass exists between 10 and 30$M_{\odot}$. Above 30$M_{\odot}$ the inferred abundance seems to be insensitive to the mass, likely because, for such large masses, the Einstein radius is very much larger than the size of the source, which behaves as point-like. In principle, if all the BHs were of 10$M_{\odot}$, the upper limit for the abundance of BHs would increase a 20\%, just to an abundance of less than 1.1\% (at 90\% confidence interval). Therefore, considering a more realistic continuous mass function for the BHs (with masses ranging between these limits) would not produce a significantly different result for the upper limit in form of BHs. A study considering the mass function of BHs (and perhaps, also of the stars) can be attempted in a future work, when more information about the BHs mass spectrum (specially at the low mass end) and light curves of many lens systems (obtained in upcoming surveys, e.g. LSST), become available. \\

Regarding the size, $r_s$, of the quasar source with which the magnification maps are convolved, it is worth mentioning the degeneracy between the microlensing based estimates of the abundance of any type of microlenses, $\alpha_{compact}$, and the adopted value of $r_s$ (Mediavilla et al. 2009). This degeneracy can be broken by anchoring the microlensing estimates of $r_s$ to reverberation mapping measurements (see Mediavilla et al. 2017 and references therein) to obtain a reference value, $r_s=5$ light-days. In principle, an increase of this parameter would result in a deeper washing-out of the imprints of the star population in the magnification maps, and, according to the $\alpha-r_s$ degeneracy, in an increase of $\alpha_{BH}$. However, unrealistic values for the quasar disk size have to be considered to obtain a significant increase (this aspect has been addressed in our accompanying letter). To show this explicitly, we have repeated the analysis for source sizes of 2.5 and 7.5 light-days. The results, presented in Figure \ref{rs_source} alongside the original result for 5 light-days, show that the PDF for the abundance of BHs is virtually unaltered, and thus, our main result (i.e. the low abundance of BHs) holds even for a 50\% of uncertainty in the adopted source size. In fact, the upper limits to the BHs abundances are $\alpha_{BH} < 0.005$ at 68\% of confidence ($\alpha_{BH} < 0.012$ at 90\%) for $r_s=2.5$ light-days and $\alpha_{BH} < 0.005$ at 68\% of confidence ($\alpha_{BH} < 0.009$ at 90\%) for $r_s=7.5$ light-days. \\

\begin{figure}[ht]
\begin{tabular}{c}
     \hskip -1.5 truecm
     \includegraphics[scale=0.75]{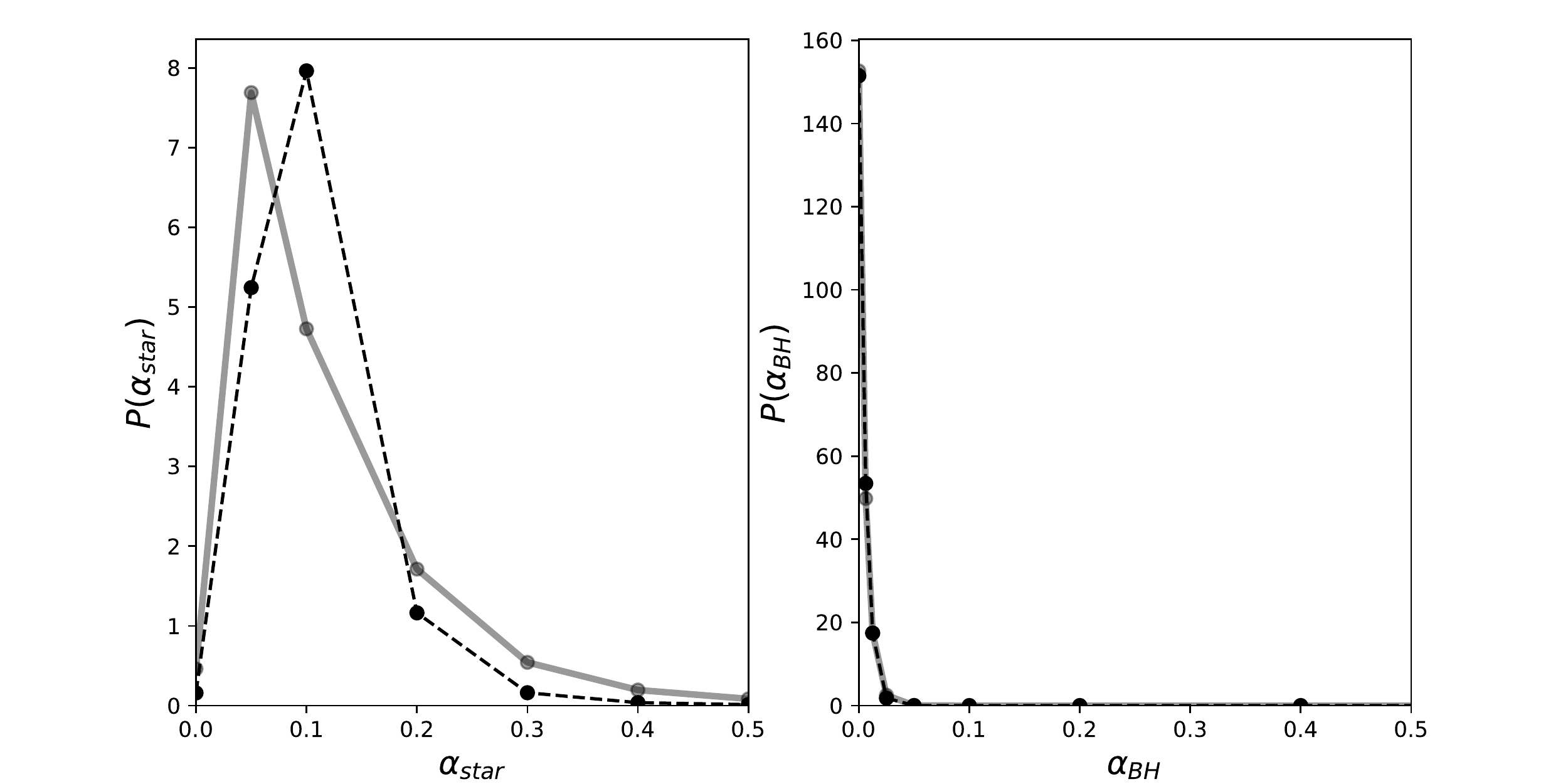} \\
     \hskip -1.5 truecm
     \includegraphics[scale=0.75]{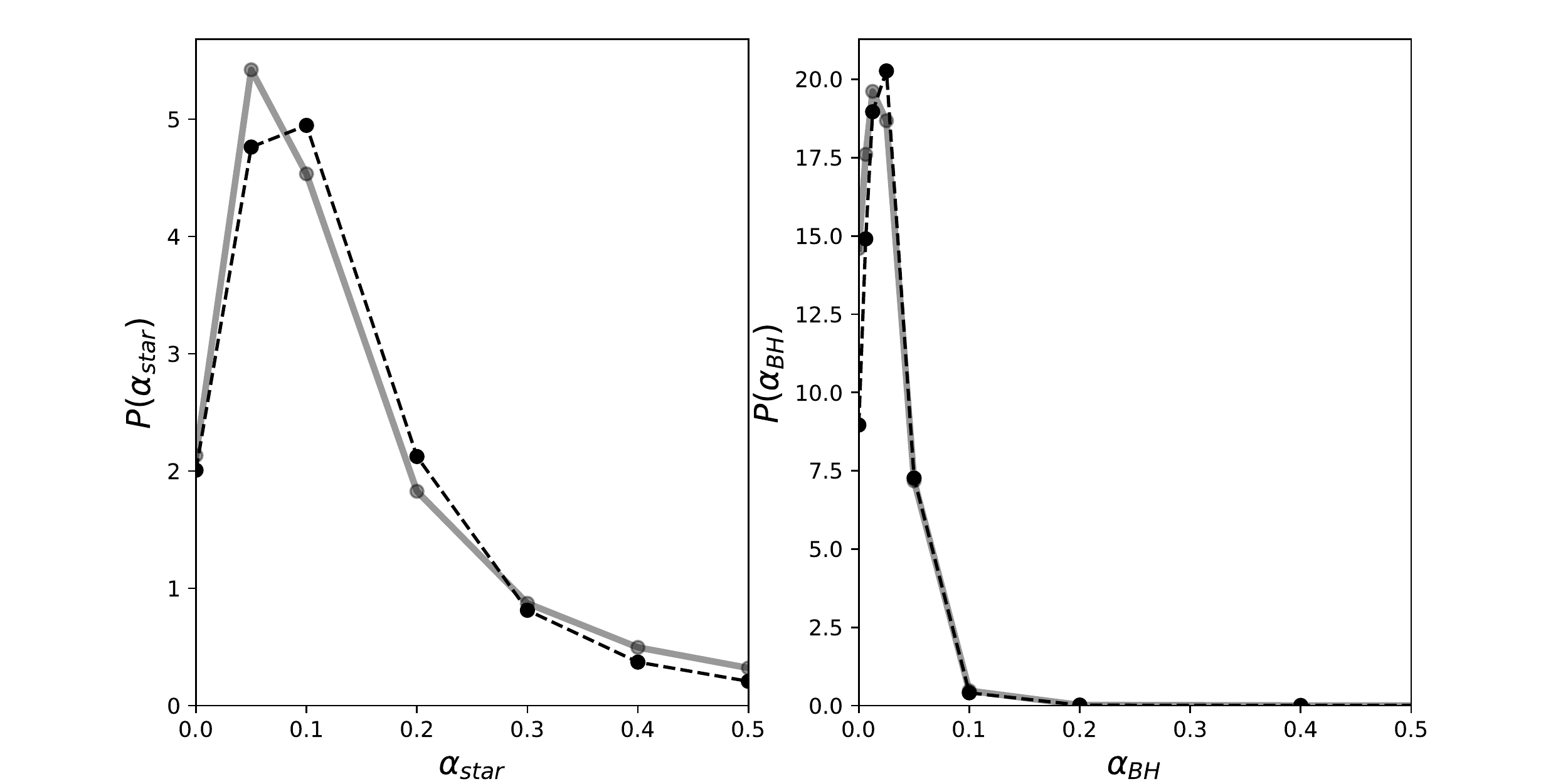} 
\end{tabular}

\caption{Dashed lines correspond to the marginalized (1D) PDFs of the abundance of stars, $\alpha_{stars}$, and BHs, $\alpha_{BH}$, obtained applying our Bayesian analysis to mock data generated supposing $\alpha_{BH}=0$ (upper panels) and $\alpha_{BH}=0.025$ (lower panels) (see text). The solid grey lines show the same PDFs marginalized over source size (see Section 4).\label{1d_mock}}
\end{figure}

The PDF marginalized over source size is included in Figure \ref{rs_source} (and also in Figure \ref{corner_plot_med}). For consistence, we have repeated this analysis and obtained the marginalized (in size) PDFs for the Bayesian analysis of mock data (see Figure \ref{1d_mock}) and for the Pooley et al. (2007) sample (see Figure \ref{corner_plot_pooley}). As commented above, this marginalization does not affect at all to the main result on the abundance of BHs. As for the stars (Figures \ref{corner_plot_med} and \ref{corner_plot_pooley}), there is a slight increase of the likelihood for small abundances (maximum now at 0.1) and a gentler decrease of the PDF for large values producing expected abundances ($\alpha_{stars}=0.18_{-0.11}^{+0.07}$ at 68\% confidence level and $\alpha_{stars}=0.18_{-0.13}^{+0.18}$ at 90\% in the case of Mediavilla et al. 2009  extended sample and $\alpha_{stars}=0.19_{-0.15}^{+0.08}$ at 68\% confidence level and $\alpha_{stars}=0.19_{-0.17}^{+0.20}$ at 90\% in the case of Pooley et al. 2007 sample) which are fully consistent with previous results (Mediavilla et al. 2017). \\

For the mass of the stars, we have chosen $m_{stars}=0.2M_{\odot}$. We take this value from Jim\'enez-Vicente et al. (2019). Although this value has also been inferred from microlensing data and might not be independent of the presence of BHs, note that any PDMF (Salpeter, Kroupa, etc.) of reasonable old ages of a few Gyr leads to a mean mass very close to this value. \\

Regarding the lens model (a SIS+$\gamma_e$ model), the $\kappa$ and $\gamma$ values corresponding to each lensed image are also subject to uncertainties, which can affect the PDFs of microlensing magnification (Vernardos \& Fluke 2014). In addition, as mentioned above, to reduce the computation time, we have generated magnification maps in only 9 bins of values of $\kappa$ and $\gamma$. Vernardos \& Fluke (2014) show that for most of the parameter space ($\kappa_{eff}$, $\gamma_{eff}$) it is reasonably to use a representative model for a nearby region in the parameter space\footnote{Particularly when considering the average values for maps smoothed to a reasonable source size, as we are basically doing in the present work (cf. their Figure 10).}. In any case, to assess the impact of these approximations, we have applied a limited bootstrapping technique, repeating all the calculations but now randomly reordering the lens models (i.e. $\kappa_i$ and $\gamma_i$) among the images in the sample\footnote{The peculiar D image of SDSS 0924+0219 has been excluded from the bootstrapping because it is very anomalous and needs to be separately modelled (see Section 3).}. We limit the random reordering to model parameters obtained adding to the fiducial ones a normal random variable of mean 0 and $\sigma$ the uncertainty in the parameters. At each iteration, we take from our discrete grid the model closest to the one randomly generated. We adopt an uncertainty of 0.035 in either $\kappa$ or $\gamma$ as typical error estimate from the thorough study by Shajib et al. (2019). We repeat the random sorting 100 times. The resulting average of the 100 random realizations shows no significant deviation from the original one. In Figures \ref{corner_plot_med} and \ref{corner_plot_pooley} we show the standard deviation of the random realizations with respect to the average\footnote{Note that the standard deviations of the bootstrapping shown in Figures 2 and 3 are not the actual errors in the PDFs, but very conservative upper limits.}. In other words, random uncertainties in the determination of the macrolens models have not a significant impact in the upper limits inferred for the BHs abundance when a large number of measurements are jointly analyzed. Notice that this analysis is equivalent to bootstrap the observed values of microlensing magnifications among different image-pairs, hence the results are also insensitive to non systematic uncertainties in the microlensing measurements. \\

The relatively low impact of bootstrapping on the results can also be explained inspecting Figure \ref{34_plots}. Microlensing measurements typically fall in a magnitude interval, for which probabilities of the $\alpha_{BH}=0$ case are greater than those of the $\alpha_{BH}=0.025$ one. Thus, as far as the random reordering of the models (or, equivalently, of the microlensing measurements) does not change the relative likelihood of the hypothesis, the procedure is rather insensitive to the random reordering. This supports the robustness of our results. \\

The determination of the no microlensing baseline (inherent to microlensing measurements, see Mediavilla et al. 2009) is, however, a source of uncertainty that may systematically affect the amplitude of microlensing magnifications.  In the optical data used by us, this baseline is determined taken as reference the broad line region, which is large enough as to be mostly insensitive to microlensing from BHs of the masses considered here. However, if these BHs were grouped in clusters, those clusters may act like pseudo-particles of a very large mass and the determination of the baseline should be revised (Heydenreich et al. 2022 in preparation). Anyhow, other microlensing magnification estimates obtained using as reference, infrared or radio data, macrolens models or light-curve monitoring, indicate that the typical amplitudes of microlensing are in the range of the optical data used by us. Indeed, as mentioned above, we have repeated the calculations for the sample of Pooley et al. (2007),  with a model based baseline, finding very similar results. The impact of clustering seems, therefore, limited. On the other hand, clusters should produce strong effects alike to millilensing, like the splitting of the lensed images, which has not been observed. \\

Our results are in good agreement with previous qualitative, indirect studies of the abundance of BHs from quasar microlensing. Mediavilla et al. (2017) obtained that the observed microlensing could be explained by a single population of stars, without considering any  contribution from intermediate mass BHs, and Esteban-Guti{\'e}rrez et al. (2020), reinterpret the results by Mediavilla et al. (2017) concluding that the existence of a significant population of BHs of intermediate mass mixed with the stars will have a low probability. Now, thanks to the present direct study, we are able to give a quantitative limit to the abundance of BHs which confirms these expectations. \\

In summary,  according to quasar microlensing data, the population of BHs with masses in the range inferred by LIGO/Virgo observations ($\sim 10M_{\odot}$ to $60M_{\odot}$), constitutes only a very small fraction of the total matter, $\alpha_{BH} \lesssim 0.004$ at 68\% of confidence ($\alpha_{BH} \lesssim 0.009$ at 90\%) for BHs masses of $\sim 30M_{\odot}$. Our present result agrees very well with the recent estimate for the fraction of mass in BHs of 0.3\% using LIGO/Virgo gravitational wave rate constraints by Wong et al. (2020). \\

\acknowledgments{We thank the anonymous referees for ideas and comments, which greatly contribute to enhance the scope of our paper. This research was supported by the Spanish MINECO with the grants AYA2016-79104-C3-1-P and AYA2016-79104-C3-3-P and projects PID2020-118687GB-C33, PID2020-118687GB-C32 and PID2020-118687GB-C31 financed by MCIN/ AEI /10.13039/501100011033. J.J.V. is supported by the project AYA2017-84897-P financed by the Spanish Ministerio de Econom\'\i a y Competividad and by the Fondo Europeo de Desarrollo Regional (FEDER), and by project FQM-108 financed by Junta de Andaluc\'\i a. AEG thanks the support from grant FPI-SO from the Spanish Ministry of Economy and
Competitiveness (MINECO) (research project SEV-2015-0548-17-4 and predoctoral
contract BES-2017-082319).}

\clearpage

\clearpage
\begin{figure*}[ht]
\hskip -0.5 truecm
\includegraphics[width =1\textwidth]{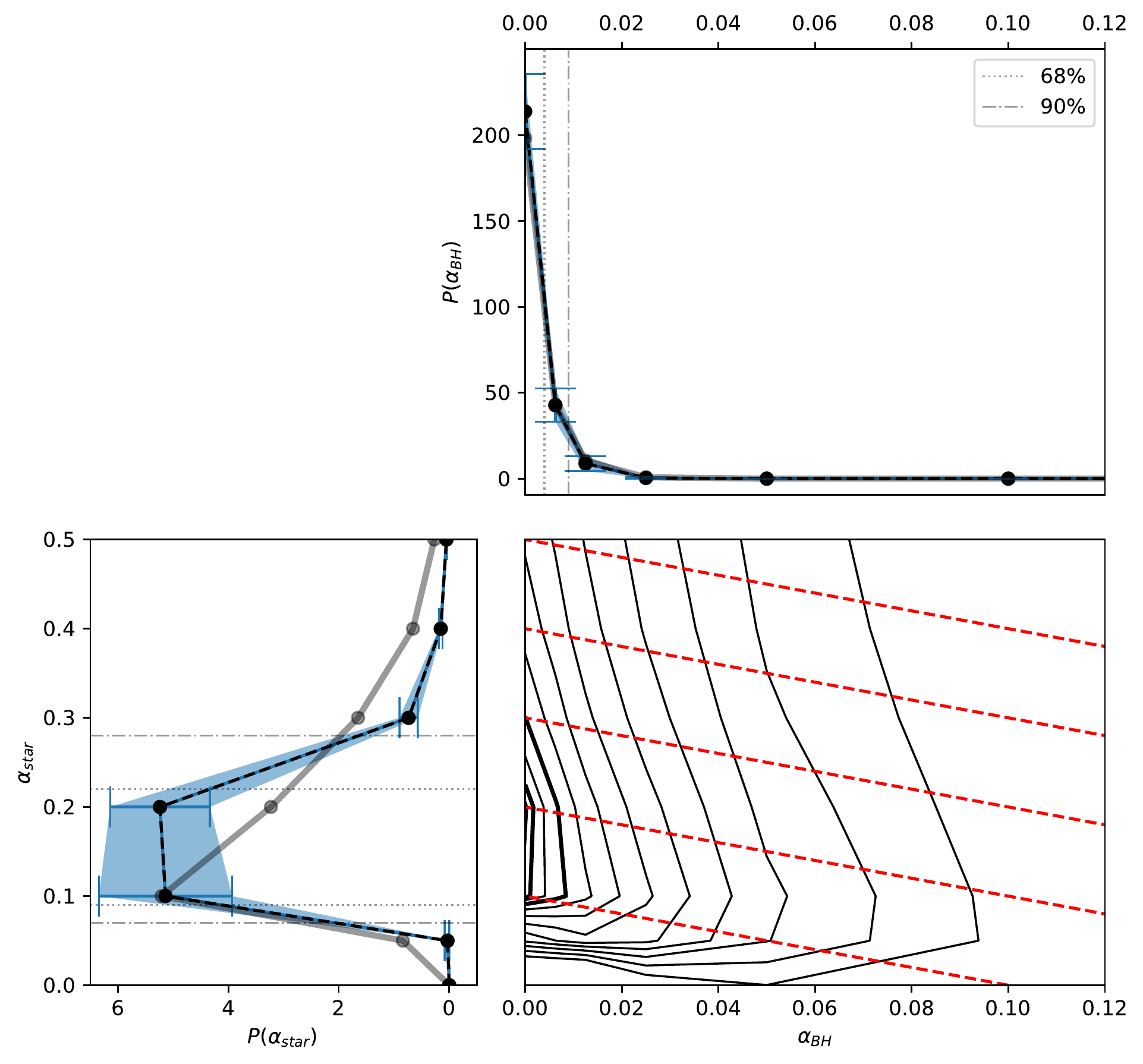}
\end{figure*}

\begin{figure*}[ht]
\caption{Probability distributions of the abundance of stars and BHs based on Mediavilla et al. (2009) extended sample. Bottom right: Joint (2D) probability density function, $p_{ijkl}({\alpha_{BH}}_k,{\alpha_{stars}}_l|\{\Delta m_{ij}\})$, of the abundance of BHs, $\alpha_{BH}$, and stars, $\alpha_{stars}$ from 1$\sigma$ to 6$\sigma$ (0.5$\sigma$ steps). Straight red dashed lines represent lines of constant $\alpha_{smooth}$ for values $\alpha_{smooth}=0.9,0.8,0.7,0.6,0.5$ (bottom to top). Thicker contours correspond to 1$\sigma$ and 2$\sigma$. Top right and bottom left: Marginalized (1D) probability density functions, $p(\alpha_{BH}|\{\Delta m_{ij}\})$ and $p(\alpha_{stars}|\{\Delta m_{ij}\})$,  of the abundance of BHs, $\alpha_{BH}$, and stars, $\alpha_{stars}$, respectively. The region $\alpha_{BH}>0.12$ (not shown) has negligible probability. Regions shaded in blue are $\pm 1$ standard deviations corresponding to lens model uncertainties estimated from a limited bootstrapping analysis. The solid grey lines show the same PDFs marginalized over source size (see Section 4). \label{corner_plot_med}}
\end{figure*}

\clearpage
\begin{figure*}[ht]
\hskip -0.5 truecm
\includegraphics[width =1\textwidth]{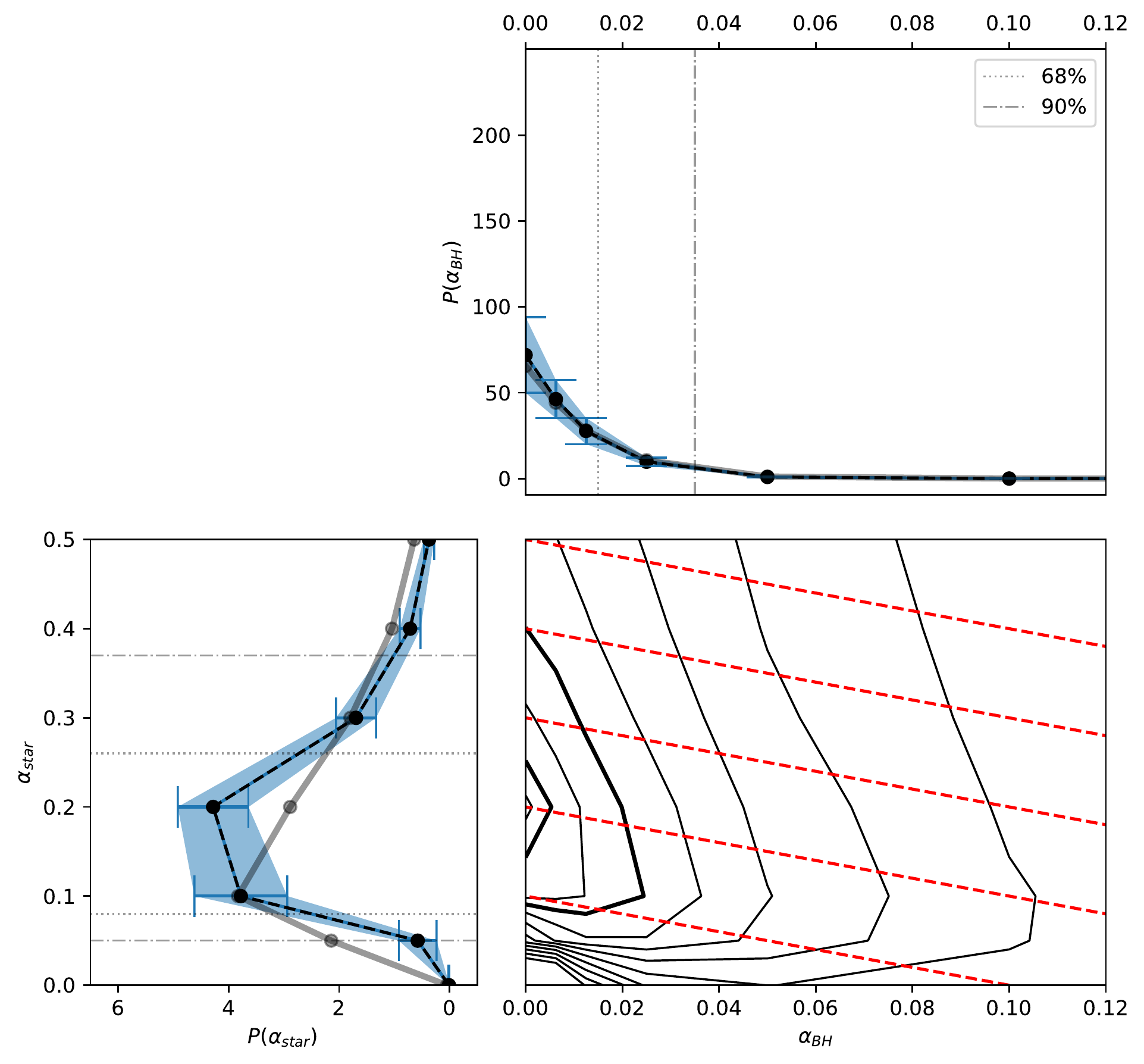}
\end{figure*}

\begin{figure*}[ht]
\caption{As Figure 2 but for Pooley et al. (2007) data. Bottom right: Joint (2D) probability density function, $p_{ijkl}({\alpha_{BH}}_k,{\alpha_{stars}}_l|\{\Delta m_{ij}\})$, of the abundance of BHs, $\alpha_{BH}$, and stars, $\alpha_{stars}$ from 1$\sigma$ to 4$\sigma$ (0.5$\sigma$ steps). Straight red dashed lines represent lines of constant $\alpha_{smooth}$ for values $\alpha_{smooth}=0.9,0.8,0.7,0.6,0.5$ (bottom to top). Thicker contours correspond to 1$\sigma$ and 2$\sigma$. Top right and bottom left: Marginalized (1D) probability density functions, $p(\alpha_{BH}|\{\Delta m_{ij}\})$ and $p(\alpha_{stars}|\{\Delta m_{ij}\})$,  of the abundance of BHs, $\alpha_{BH}$, and stars, $\alpha_{stars}$, respectively. The region $\alpha_{BH}>0.12$ (not shown) has negligible probability. Regions shaded in blue are $\pm 1$ standard deviations corresponding to lens model uncertainties estimated from a limited bootstrapping analysis. The solid grey lines show the same PDFs marginalized over source size (see Section 4). \label{corner_plot_pooley}}
\end{figure*}

\clearpage
\begin{figure}[ht]
\hskip -1.5 truecm
\includegraphics[scale=0.7]{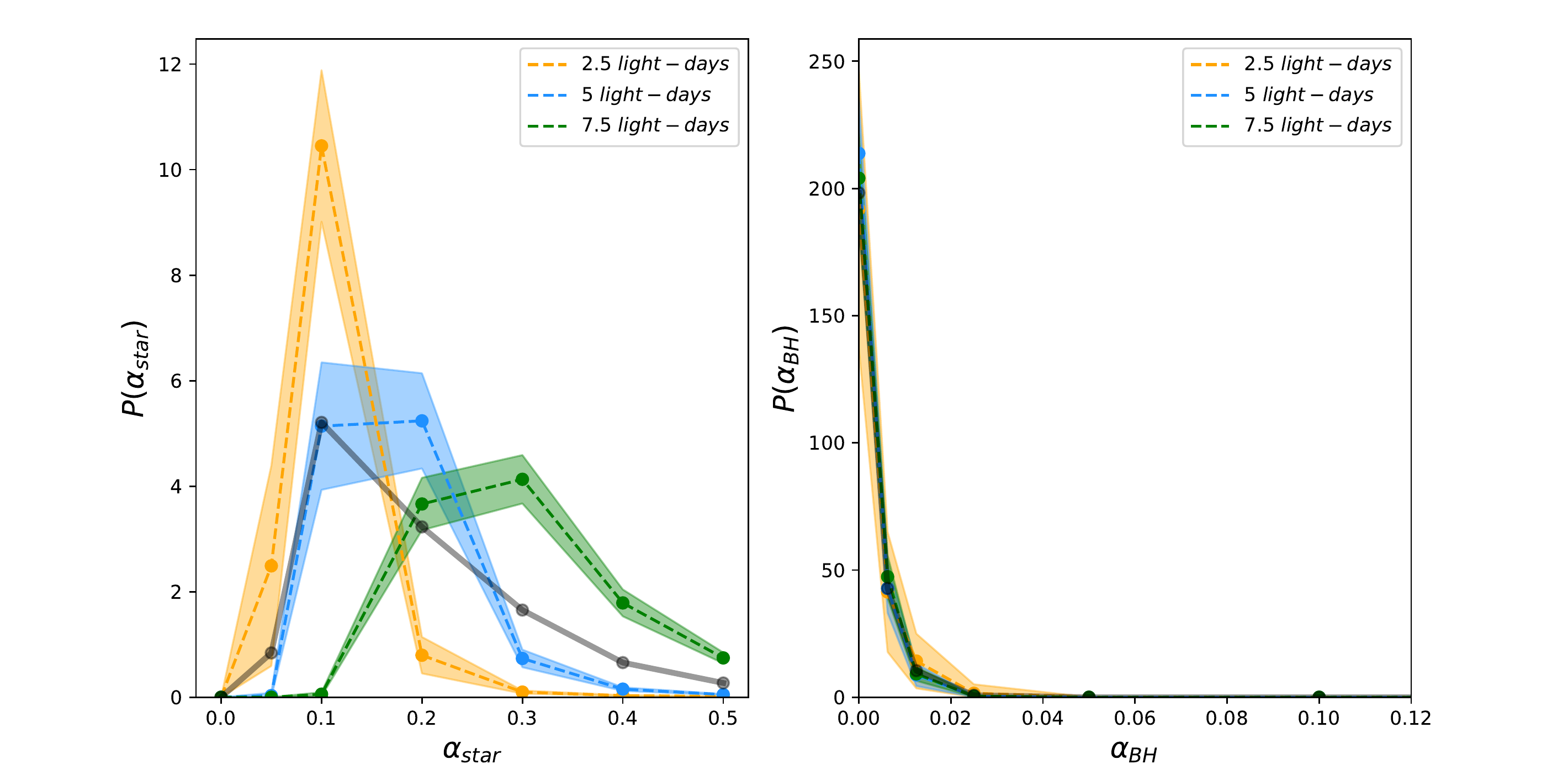}
\caption{Marginalized (1D) PDFs of the abundance of stars, $\alpha_{stars}$, and BHs, $\alpha_{BH}$, for three different source sizes $r_{s}=2.5, 5, 7.5$ light-days (orange, blue and green, respectively) obtained from Mediavilla et al. (2009) extended sample (see text). The region $\alpha_{BH}>0.12$ (not shown) has negligible probability. Shaded regions show $\pm 1$ standard deviations corresponding to lens model uncertainties estimated from a limited bootstrapping analysis. The solid grey lines show the same PDFs marginalized over source size (see Section 4).\label{rs_source}}
\end{figure}

\vspace{10mm}

\begin{deluxetable}{ccccccc}[ht]
\tablenum{1}
\tablewidth{1pt}
\tablecaption{Measurements of microlensing magnifications \label{tab:deltam}}
\tablehead{\colhead{Object} & \colhead{Image pair} & \colhead{$\Delta m_{ij}$} & \colhead{$\kappa_{1}$} & \colhead{$\kappa_{2}$} & \colhead{$\gamma_{1}$} & \colhead{$\gamma_{2}$}}
\vspace{-0.5cm}
\startdata
HE 0047-1756$^{1}$ & $B-A$ & -0.6 & 0.43 & 0.61 & 0.48 & 0.65 \\
HE 0435-1223$^{2}$ & $D-B$ & 0.26 & 0.52 & 0.56 & 0.59 & 0.64 \\
SDSS J0924+0219$^{3}$ & $C-B$ & 0.27, 0.66, 0.29 & 0.45 & 0.57 & 0.39 & 0.59 \\
QSO 0957+561$^{4}$ & $B-A$ & -0.44 & 0.20 & 1.03 & 0.15 & 0.91 \\
SDSS J1004+4112$^{4}$ & $B-A$ & -0.40 & 0.48 & 0.48 & 0.59 & 0.48 \\
HE 1104-1805$^{4}$ & $B-A$ & 0.56, 0.07 & 0.64 & 0.33 & 0.52 & 0.21 \\
WFI J2033-4723$^{2}$ & $C-B$ & 0.31, 0.48 & 0.38 & 0.61 & 0.25 & 0.73 \\
QSO 1355-2257$^{3}$ & $B-A$ & 0.41, 0.47 & 0.30 & 1.10 & 0.29 & 1.08 \\
SDSS1029+2623$^{3,4}$ & $B-A$ & 0.008, 0.40 & 0.57 & 0.52 & 0.30 & 0.40 \\
HE2149+2745$^{2}$ & $B-A$ & 0.23 & 0.31 & 1.25 & 0.32 & 1.25 \\
SDSS1155+6346$^{1}$ & $B-A$ & -0.58 & 0.22 & 1.67 & 0.03 & 1.47 \\
\enddata
\tablecomments{References. $^{1}$ Rojas et al. (2014). $^{2}$ Motta et al. (2017). $^{3}$ Rojas et al. (2020). $^{4}$ Motta et al. (2012).}
\end{deluxetable}
\clearpage

\appendix
\renewcommand\thefigure{\thesection.\arabic{figure}}
\section{Sensibility of the method to the presence of Black Holes}
To explain the strong constraint imposed by our analysis on the abundance of BHs, in Figure \ref{34_plots} we plot the $\alpha_{BH}=0$ and $\alpha_{BH}=0.025$ PDFs (for $\alpha_{stars}=0.1$) for the 34 image pairs considered. We also mark the observed microlensing magnification difference for each pair (a total of 44 measurements). As it can be readily seen in Figure \ref{34_plots}, for the measured microlensing, the probability of the $\alpha_{BH}=0$ PDF is in almost all the cases significantly greater than the probability of the $\alpha_{BH}=0.025$ one. This explains the high probability of $\alpha_{BH}=0$ when the 44 measurements are jointly considered. \\

\setcounter{figure}{0}
\begin{figure}[ht]
\hskip -1 truecm
\includegraphics[scale=0.4]{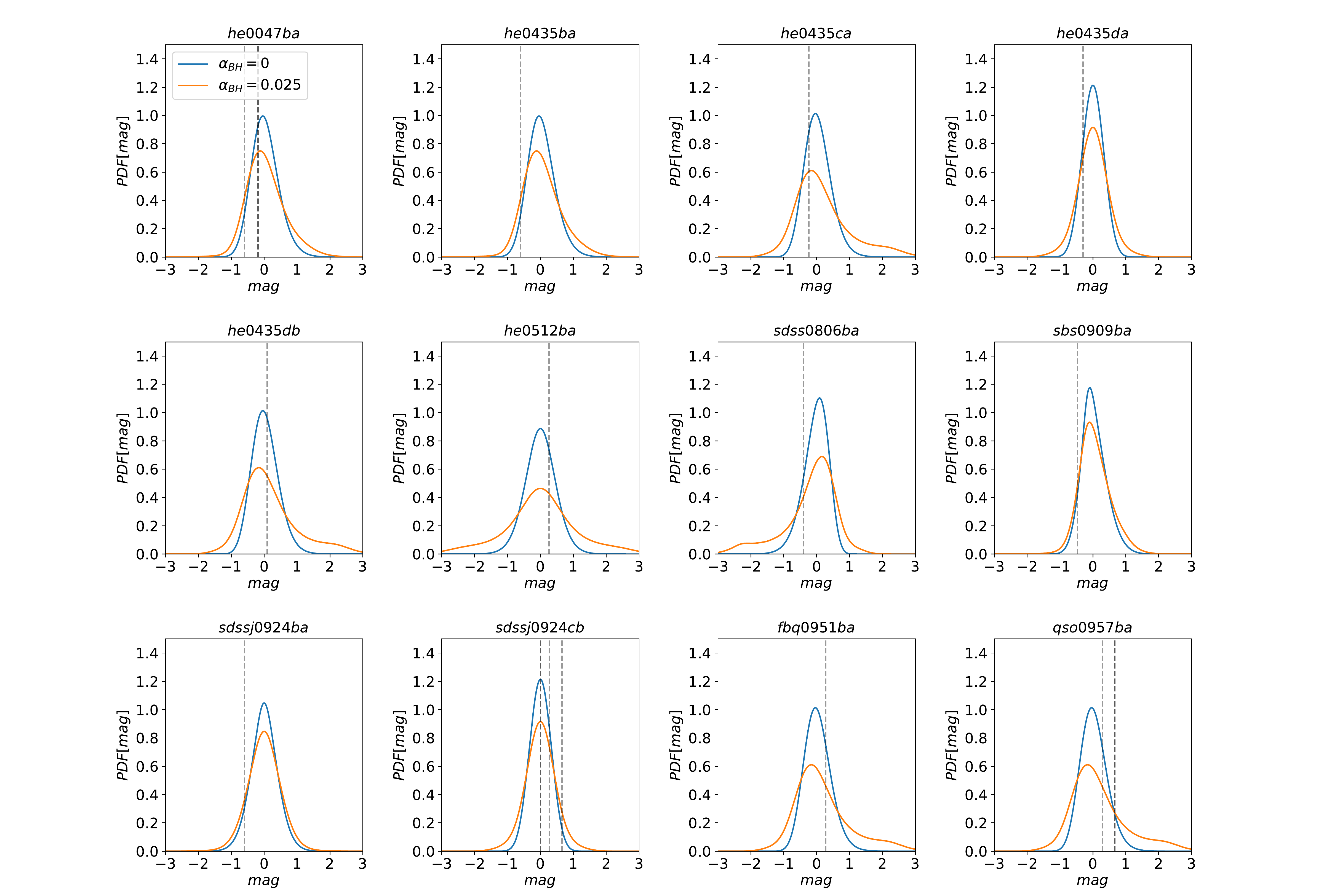}
\end{figure}

\begin{figure}[ht]
\hskip -1 truecm
\includegraphics[scale=0.4]{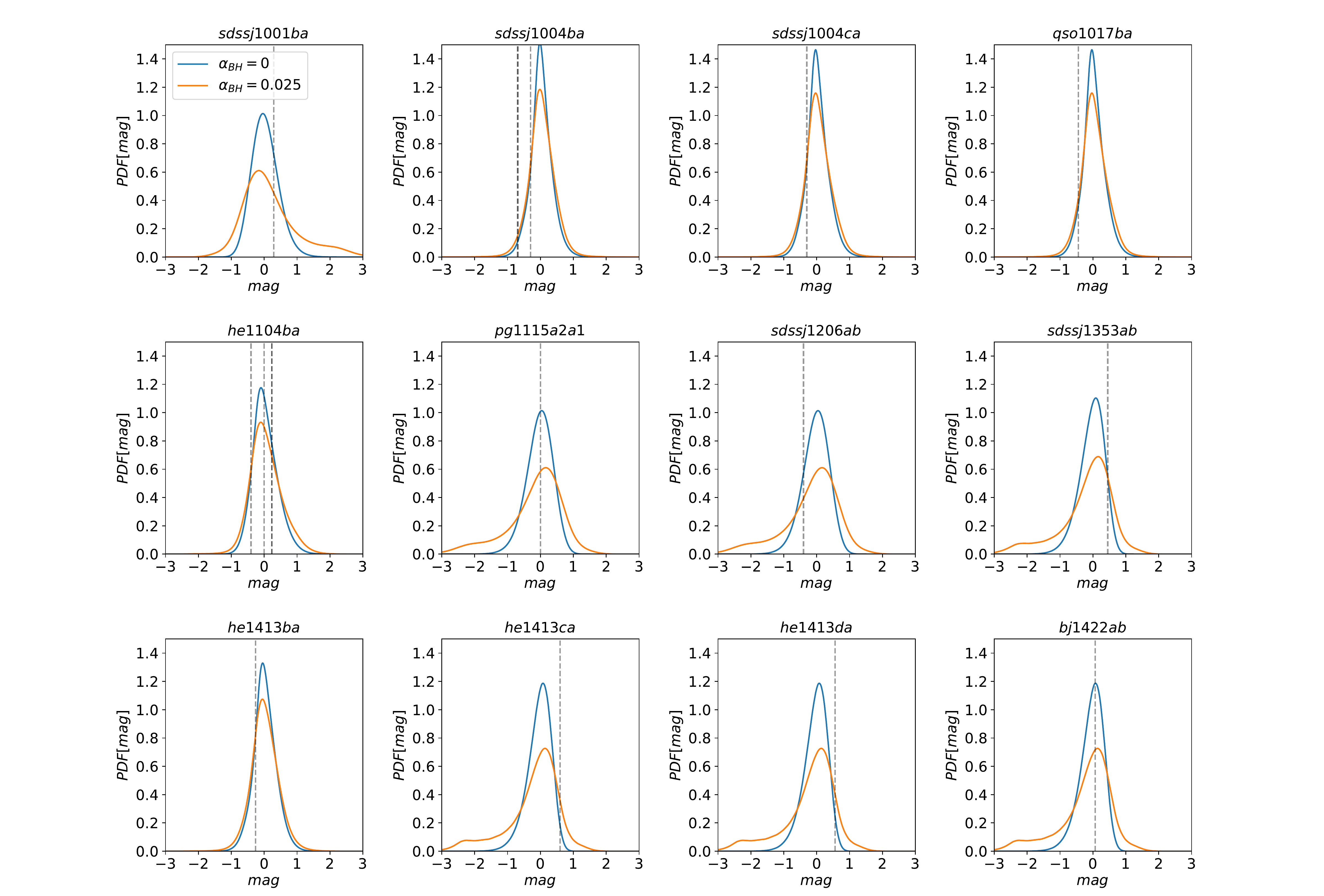}
\end{figure}

\begin{figure}[ht]
\hskip -1 truecm
\includegraphics[scale=0.4]{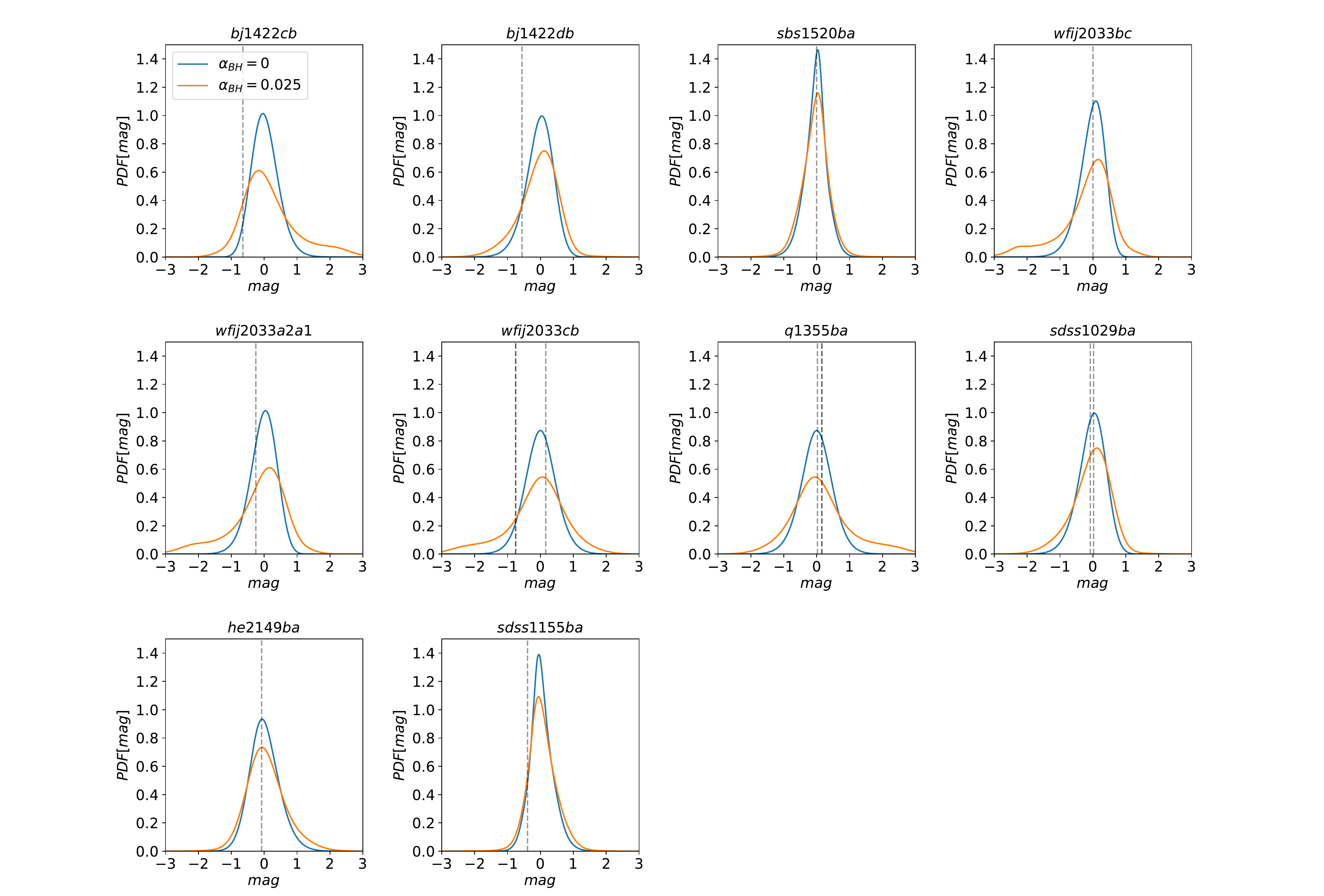}

\caption{Probability density functions (PDFs) for the $\alpha_{BH} = 0$ (blue) and $\alpha_{BH}=0.025$ (orange) cases with $\alpha_{stars}=0.1$. Each panel corresponds to one of the 34 image pairs considered (see text). The vertical dashed lines mark measured differential microlensing (one or more) for the image pair (a total of 44 measurements).\label{34_plots}}
\end{figure}

\end{document}